\documentclass[reprint,aip,jcp,showpacs,superscriptaddress,preprintnumbers,amsmath,amssymb,floatfix]{revtex4-1}
\usepackage{graphicx}
\usepackage{tabularx}
\usepackage{dcolumn}
\usepackage{amsmath}
\usepackage{amssymb}
\usepackage{bm}
\usepackage{graphics}
\usepackage{subfigure}
\usepackage{booktabs}
\usepackage{array}
\usepackage{xcolor}
\begin{document}

\newcommand{\arhplus}{\ensuremath{\mathrm ArH^+}}
\newcommand{\htwo}{\ensuremath{\mathrm H_2}}
\newcommand{\htwoplus}{\ensuremath{\mathrm H_2^+}}
\newcommand{\hhh}{\ensuremath{\mathrm H_3^+}}

\title{Visible line intensities of the triatomic hydrogen ion from experiment and theory}

\author{Annemieke Petrignani}%
\altaffiliation[Present address: ]{Leiden Observatory, Leiden University, PO Box  9513, NL-2300 RA Leiden, The Netherlands}
\author{Max H. Berg}%
\author{Florian Grussie}%
\author{Andreas Wolf}%
\affiliation{Max-Planck-Institut f\"{u}r Kernphysik, Saupfercheckweg 1, D-69117 Heidelberg, Germany}
\author{Irina I.  Mizus}%
\affiliation{Institute of Applied Physics, Russian Academy of Science, Ulyanov 
Street 46, Nizhnii Novgorod, Russia 603950}%
\author{Oleg L. Polyansky}%
\affiliation{Institute of Applied Physics, Russian Academy of Science, Ulyanov 
Street 46, Nizhnii Novgorod, Russia 603950}%
\affiliation{Department of Physics and Astronomy, University College London, 
London WC1E 6BT, UK}%
\author{Jonathan Tennyson}%
\affiliation{Department of Physics and Astronomy, University College London, 
London WC1E 6BT, UK}%
\author{Nikolai F. Zobov}%
\affiliation{Institute of Applied Physics, Russian Academy of Science, Ulyanov 
Street 46, Nizhnii Novgorod, Russia 603950}%
\author{Michele Pavanello}%
\affiliation{Department of Chemistry, Rutgers University, Newark, NJ 07102, USA}%
\author{Ludwik Adamowicz}%
\affiliation{Department of Chemistry and Biochemistry, The University of Arizona, Tucson AZ 
85721, USA}%

\date{\today}

\begin{abstract}
The visible spectrum of \hhh\ is studied using high-sensitivity action spectroscopy in a cryogenic radiofrequency multipole trap. Advances are made to measure the weak ro-vibrational transitions from the lowest rotational states of \hhh\ up to high excitation energies providing visible line intensities and, after normalisation to an infrared calibration line, the corresponding Einstein $B$ coefficients. {\it Ab initio} predictions for the Einstein $B$ coefficients are obtained from a highly precise dipole moment surface of \hhh\ and found to be in excellent agreement, even in the region where states have been classified as chaotic.
\end{abstract}

\maketitle

The triatomic molecular ion \hhh\ prevails in cold hydrogen plasmas and has therefore been the subject of
study for many decades in astronomy and for even longer in molecular physics. In space, \hhh\ is the cornerstone of the chemical network leading to larger molecules \cite{jt157} and it is now being widely monitored in both the dense \cite{mccall:1999} and diffuse \cite{12InMc} interstellar medium. In planetary atmospheres, it acts as a tracer of \htwo, as an important coolant, and as a tracer of ionospheric activity. \cite{jt258}  \hhh\ is thought to be the key of the thermal stability of hot Jupiter exoplanets. \cite{koskinen:2007} For modeling the role of \hhh\ in these environments and for interpreting related observations, precise and reliable data on the \hhh\ spectrum are required in an increasingly wide wavelength range. Such work relies heavily on comprehensive, theoretical spectral line calculations including transition energies and also line intensities. \cite{jt107,jt181} While theoretical line intensities in comparison to laboratory observations served as a guide in the assignment of complex discharge spectra in the infrared (IR) spectral region, \cite{lindsay:2001a} experimental information on the intensities for independently identified lines is scarce.

By combining laser spectroscopy in a cryogenic trap with high-precision calculations of the ground-state potential energy surface of the ion, we recently identified several lines of \hhh\ in the near-visible and visible (VIS) spectral range and showed that our theoretical transition energies agreed with the measured line positions on the sub-tenth cm$^{-1}$ level. \cite{jt512} Here we prepare \hhh\ ions in their lowest {\em ortho} and {\em para} rotational levels only and measure the intensities of eighteen weak near-VIS and VIS spectral lines, which are directly referred to the intensity of an IR calibration transition whose value can be safely predicted. Theoretically, we developed a dipole moment surface (DMS) for \hhh\ that also covers the VIS part of its vibrational spectrum, which we used to model the line intensity measurements. 

\begin{table*}[t]
  \caption{Observed near-VIS and VIS transitions of H$_3^+$ with their assignments, measured transition wave number $\bar{\nu}_m$,
    and their intensity relative to the IR reference line. The transitions start from the vibrational ground state $00^0$. 
    \label{tab1:intensities}}
    \begin{ruledtabular}
    \begin{tabular*}{\textwidth}{clllddd}
      & & \multicolumn{1}{c}{$\bar{\nu}_m$(cm$^{-1}$)} & \multicolumn{1}{c}{Intensity ratio} & \multicolumn{3}{c}{$B$ ($10^{18}$ cm$^3$\,J$^{-1}$\,s$^{-2}$)} \\
      $\Delta J(J'',G'')$ & \multicolumn{1}{c}{$v'_1{v'_2}^{\left|l'\right|}(J',G')$}  & 
      \multicolumn{2}{c}{This experiment} & \multicolumn{1}{c}{This exp.\ %
        \footnote{Using $B = 1.54(2)\times10^{21}$ cm$^3$\,J$^{-1}$\,s$^{-2}$ (this theory) for the reference line 03$^1$ $\leftarrow$
          $00^0$ $R(1,1)^l$ at 7144.235 cm$^{-1}$}} & 
      \multicolumn{1}{c}{This theory} & \multicolumn{1}{c}{NMT\footnote{Reference \citenum{jt181}}} \\ 
      \hline\\[-2ex]
      $R(1,0)$  	& $22^2\,(2,3)$ 	&  10\,752.161(5)	& 0.047(10)		& \multicolumn{1}{c}{~ 73(16)}	
																&  75.9(78)	& 60.5 	\\
      $P(1,1)$  	& $05^1\,(0,1)$ 	&  10\,798.777(5)	& 0.017\,2(19)		& 26.5(30) &  34.6(12)	& 32.9	\\
      $Q(1,0)$  	& $05^1\,(1,0)$ 	&  10\,831.681(5)	& 0.073(10)		& \multicolumn{1}{c}{~112(16)}  
																&  98.2(37)   	& 94.0 	\\
      $Q(1,1)$  	& $23^1\,(1,1)$ 	&  12\,373.526(10)	& 0.002\,77(64)		& 4.3(10)	&   4.87(83)	&  4.04  	\\
      $R(1,1)$ 	& $23^1\,(2,1)$
                                			&  12\,381.137(10) 	& 0.002\,68(66)  	& 4.1(10)	&   4.67(52)	&  4.10  	\\
      $P(1,1)$  	& $06^2\,(0,2)$ 	&  12\,413.247(10)	& 0.002\,98(75)		& 4.6(12)	&   3.85(28)  	&  3.73  	\\
      $R(1,1)$  	& $23^1\,(2,1)$
                                			&  12\,588.951(5) 	& 0.000\,72(25)  	& 1.10(38)	&   1.05(18)	&  0.86 	\\
      $R(1,1)$  	& $23^3\,(2,5)$ 	&  12\,620.223(5)	& 0.004\,07(80)		& 6.3(12)	&   5.09(31)  	&  4.86 	 \\
      $R(1,1)$  	& $06^2\,(2,4)$ 	&  12\,678.683(10)	& 0.005\,5(11)		& 8.6(17)	&   8.74(49)  	&  8.01  	\\
      $R(1,1)$  	& $32^2\,(2,4)$ 	&  13\,332.884(10)	& 0.002\,61(84)  	& 4.0(13)	&   2.86(40)  	&  2.05 	\\
      $P(1,1)$  	& $07^1\,(0,1)$ 	&  13\,638.251(5)	& 0.002\,54(95)  	& 3.9(15)	&   4.47(44)  	&  4.14 	\\
      $P(1,1)$ 	& $08^2\,(0,2)$ 	&  15\,058.680(5)	& 0.001\,00(22)  	& 1.53(33)	&   1.76(27) 	&  1.62 	\\
      $Q(1,1)$  	& $16^2\,(1,2)$ 	&  15\,130.480(5)	& 0.000\,47(10)  	& 0.72(16)	&   0.83(22) 	&  0.71 	\\
      $R(1,0)$  	& $16^2\,(2,3)$ 	&  15\,450.112(5)	& 0.000\,489(66) 	& 0.75(10)	&   0.92(31) 	&  0.78 	\\
      $R(1,0)$  	& $16^4\,(2,3)$ 	&  15\,643.052(5)	& 0.000\,718(98) 	& 1.11(15)	&   1.18(30) 	&  1.01 	\\
      $Q(1,1)$  	& $25^1\,(1,1)$ 	&  15\,716.813(5)	& 0.001\,04(33)  	& 1.60(51) &   1.56(24) 	&  1.40 	\\
      $R(1,0)$  & $34^2\,(2,3)$ 	&  16\,506.139(5) 	& 0.000\,83(32)  	& 1.28(50) &   1.23(23) 	&  		\\
      $Q(1,1)$  & $J'=1$ \footnote{Further assignment unavailable} 
						&  16\,660.240(5) 	& 0.000\,24(12)  	& 0.38(19) &   0.62(18) 	&  		\\
    \end{tabular*}
    \end{ruledtabular}
\end{table*}

Highly excited molecular motion forms a challenge in the {\it ab initio} treatment of molecules and molecular dynamics.
\cite{Morong:2009,jt582} The Born--Oppenheimer approximation breaks down \cite{jt566} and the molecular motions couple strongly to each other, leading to ill-defined approximate quantum numbers and vibrational angular momenta. \cite{Watson:1994} This behavior is associated with the onset of chaos in the analogous classical problem. Non-adiabatic effects are often only included as {\it ad hoc} correction factors on the calculations. \cite{jt236} Despite these difficulties, the recent advanced theoretical treatment \cite{jt512} gave spectroscopic accuracy in the predictions of \hhh\ ro-vibrational energy levels up to 2 eV internal excitation.

The investigation of high levels of excitation also constitutes a great experimental challenge as the spectra often become complex and unassignable. Carrington {\em et~al.} \cite{Carrington:1982,Carrington:1984} triggered the extensive theoretical investigation of \hhh\ with their measured, rich spectrum of states near the dissociation limit, which remains unassigned. The \hhh\ internal motion already becomes chaotic at about one-third of the dissociation energy, \cite{jt157} where the nuclei sample linear configurations. The first measurements just above this barrier to linearity were performed by Gottfried and coworkers using advanced absorption spectroscopy. \cite{Gottfried:2006,Gottfried:2003} This method reaches its limit at 13\,700 cm$^{-1}$, where the absorption Einstein coefficient $B$ of the transitions reaches a few $10^{18}$ cm$^3$\,J$^{-1}$\,s$^{-2}$. No Einstein coefficients could be derived due to unknown initial population of the \hhh\ levels. The identification of the many lines was also hindered by the large number of initial levels involved.

The development of action spectroscopy in a cryogenic radiofrequency (RF) ion trap provided measurements in the near-IR range \cite{Kreckel:2008} and showed great promise for attaining much higher sensitivity than absorption spectroscopy. We have recently extended the range of this method up to almost 17\,000 cm$^{-1}$. \cite{jt512} The measurement of line intensities in the VIS region has, however, been out of reach until now. As described previously \cite{jt512,Kreckel:2008,Berg:2012} \hhh\ ions were stored in a multipole RF ion trap and buffer-gas cooled with He ($2 \times 10^{14}$ cm$^{-3}$) to $\sim85$~K. About 85\% of the \hhh\ ions thus populated the two lowest rotational ($J=1$) levels of the vibrational ground state, having {\em ortho} and {\em para} symmetry. The near-VIS to VIS transitions were measured by chemical probing with Ar ($1 \times 10^{12}$ cm$^{-3}$). The reaction \hhh\ + Ar $\rightarrow$ ArH$^+$ + \htwo\ is endothermic but activated by ro-vibrational excitation of \hhh\ above 4500 cm$^{-1}$. The count of extracted ArH$^+$ reaction products per storage cycle is the spectroscopic signal and is only formed by \hhh\ ions excited above the endothermicity barrier. There is a background from non-laser induced ArH$^+$, originating from the initially hot \hhh\ ions. The decay rate of this background ArH$^+$ is determined by the back reaction of ArH$^+$ with \htwo\ ($2 \times 10^{10}$ cm$^{-3}$), while the forward reaction is stopped within a few ms by the buffer-gas cooling of \hhh. A pre-cooling time of 200 ms was chosen to achieve a sufficiently low ArH$^+$ background level of less than 0.1 counts on average per storage cycle. Afterwards, the \hhh\ ions were irradiated for 100 ms yielding, when on resonance, internal \hhh\ ion energies much higher than the required reaction barrier. The reaction time of $\sim$0.5 ms for ArH$^+$ formation, estimated from the Langevin cross section, is shorter than the effective radiative decay time of at least several ms. \cite{jt181} Collisional de-excitation of vibrations can be neglected, even for He which has the highest partial pressure. After irradiation, the ion cloud was released into a mass spectrometer (MS) that could be set to either ArH$^+$ and \hhh\ detection from one 300-ms storage cycle to the next. The ions were counted on a Daly stamp with a scintillator and a photomultiplier detection system \cite{Daly:1960,Kreckel:2008} that is sensitive enough to single-ion count the low ArH$^+$ and the high \hhh\ (normalization) ion numbers.

To determine the intensity of the spectral lines, we performed scans of the laser wavelength in small steps around the central wavelength pre-located in wider search scans. \cite{jt512} The strength of the ArH$^+$ counting signal exceeding the background level was determined from a Gaussian fit to the scanned line. The observed lines were Doppler-broadened corresponding to a kinetic temperature near 85 K. The experimental challenge for reliable intensity measurements is not only to increase the spectroscopic sensitivity to a level where we can measure these very weak, lines \cite{jt512} but also to increase the dynamical counting range for both the ArH$^+$ signal and the \hhh\ normalization signal, whilst retaining linearity and constant (within the Poisson noise) experimental parameters. To increase the dynamic range to 10$^4$, we applied a voltage ramp on the trap exit electrode to gradually release the ions, leading to a stretched ion pulse train of 0.5--1 ms and yielding a factor of 25 improvement in dynamic range. Group recordings of 10 storage cycles each were alternately performed with the MS set on \hhh\ (normalization) or on ArH$^+$ with (signal) and without (background) laser. Two spectroscopy lasers were applied along the trap axis: a single-mode Ti:sapphire or a dye laser for measuring a near-VIS or VIS line, and a single-mode IR laser diode for measuring an IR reference line, 03$^1$ $\leftarrow$ $00^0$ $R(1,1)^l$ at 7144.235 cm$^{-1}$. \cite{lindsay:2001a} Both lasers filled identical apertures, ensuring their spatial overlap inside the trapping region. A measurement on a near-VIS or VIS line typically lasted 15~h, and immediately before and after this, 2-h measurements were performed on the IR reference line. In a campaign extending over several months, intensity measurements relative to the IR reference line were performed on 18 near-VIS or VIS lines of \hhh. These intensity measurements were performed separately from the runs aiming at the line frequency determination described earlier. \cite{jt512}

The intensity ratios, linking the near-VIS to VIS lines to a single IR calibration line far below the barrier to linearity, are listed together with the measured transition wave numbers $\bar{\nu}_m$ in Table \ref{tab1:intensities}.  In the line assignments, \cite{alijah10,jt512} $J''$ and $J'$ refer to the lower and upper level, respectively.  Also listed are values of the absolute rate coefficients obtained for the Einstein coefficient of the IR calibration line from the theoretical method of the present work; previous theoretical calculations \cite{jt181} of this calibration value yielded essentially the same result ($B=1.55 \times 10^{21}$ cm$^3$\,J$^{-1}$\,s$^{-2}$).

In our calculations on the two-electron system \hhh, the spatial component of the electronic wave function $\Phi_M(\mathbf{r})$, is expanded in terms of a set of $M$ explicitly correlated Gaussians (ECGs), $\{g_k\}_{k=1,\ldots M}$, as:
\begin{equation}\label{eq:ecsg1}
\Phi_M(\mathbf{r})=\sum_{k=1}^M C_k {\cal P} g_k(\mathbf{r}),
\end{equation}
\noindent where ${\cal P}$ is the permutational-symmetry projection operator which symmetrizes each ECG with respect to the permutation of the electron labels. The ECGs are the following 6-dimensional functions:
\begin{equation}\label{eq:ecsg:2}
g_k(\mathbf{r})=\exp
\left[-(\mathbf{r}-\mathbf{s}_k)'\overline{\mathbf{A}}_k
(\mathbf{r}-\mathbf{s}_k) \right],
\end{equation}
\noindent where $\mathbf{r}$ and $\mathbf{s}_k$ are 6-dimensional vectors of the electronic Cartesian coordinates and of the coordinates of the Gaussian shifts, respectively, and the prime denotes vector transposition.  $\overline{\mathbf{A}}_k$ is a $6\times6$ symmetric matrix of the Gaussian exponential parameters defined as:
\begin{equation}
\overline{\mathbf{A}}_k = \mathbf{A}_k \otimes \mathbf{I}_3,
\end{equation}
\noindent with $\mathbf{I}_3$ being the $3\times 3$ identity matrix and $\otimes$ denoting the Kronecker product and $\mathbf{A}_k$ is a $2\times2$ symmetric matrix. More details on how the  spatial wave function is constrained to satisfy the Pauli exclusion principle, and to be square integrable are given elsewhere. \cite{jt526,bubin2012a}

\begin{figure}[t]
  \includegraphics[width=8.5cm]{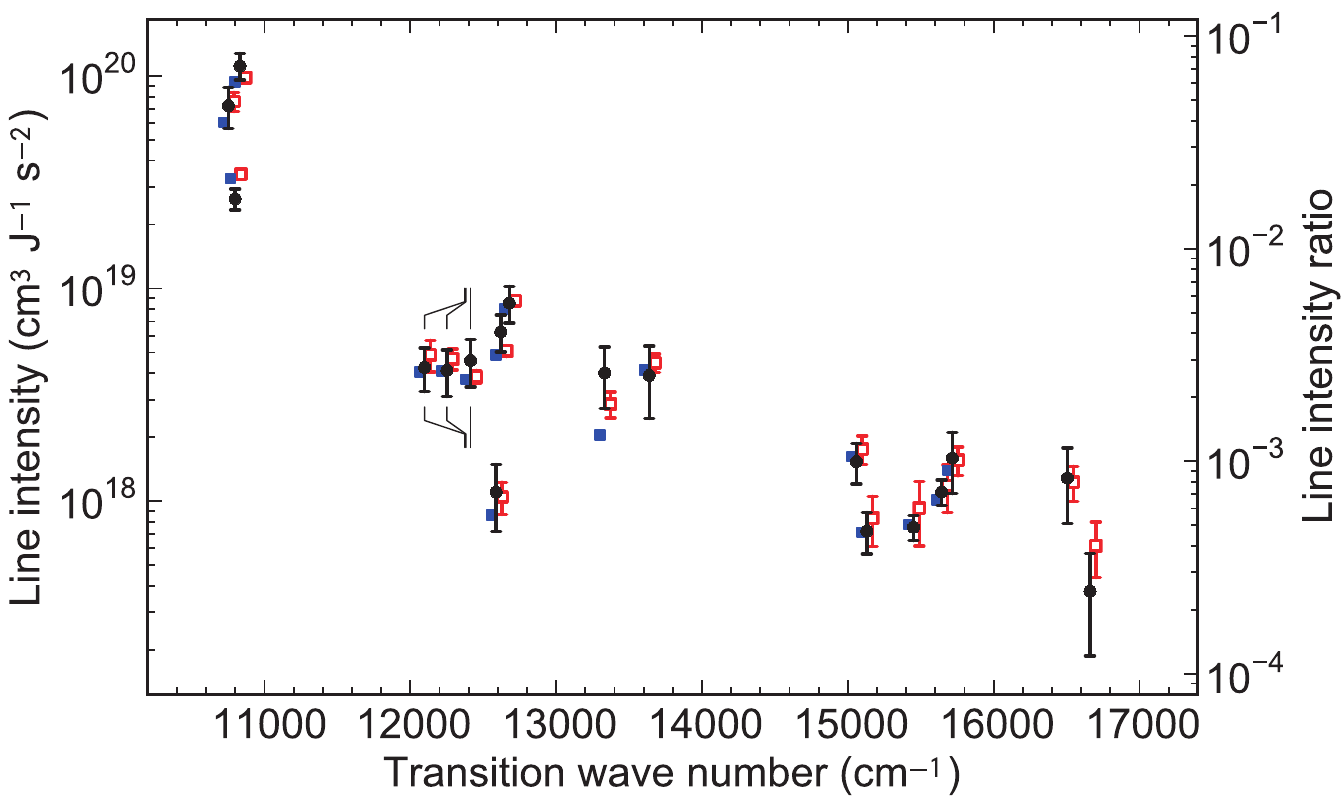}
  \caption{Line intensities for near-VIS and VIS transitions of \hhh\ as a function of the transition wave number (horizontally shifted where helpful for clarity). Experimental results (dots and error bars) derived from the measured line intensity ratios (right-hand scale) and theoretical results from this work (open squares) and from NMT \cite{jt181} (full squares, where available).
    \label{fig1:intensities}}
\end{figure}

Here calculations of the dipole moment surface (DMS) are presented. The dipole moment is determined as an expectation value of the dipole operator with the variationally optimized electronic wave function (\ref{eq:ecsg1}) containing 900 ECGs for each of roughly 42~000 geometry points. \cite{jt512,jt526} The dipole moment vector associated with a certain \hhh\ molecular configuration defined by the Cartesian position coordinates ($\mathbf{R}_1$, $\mathbf{R}_2$, and $\mathbf{R}_3$) is:
\begin{eqnarray}\label{eq:dip1}
 \boldsymbol\mu & = &
\sum_{kl}^M C_k C_l \langle {\cal P} g_k(\mathbf{r}) | \left(-\sum_{i=1}^2 \mathbf{r}_i \right)
| {\cal P} g_l(\mathbf{r}) \rangle + \sum_{\alpha=1}^3 \mathbf{R}_\alpha \nonumber \\ 
& = & \sum_{kl}^M C_k C_l \langle g_k(\mathbf{r}) | \left(-\sum_{i=1}^2 \mathbf{r}_i \right)| 
{\cal P} g_l(\mathbf{r}) \rangle + \sum_{\alpha=1}^3 \mathbf{R}_\alpha,
\label{dipole}
\end{eqnarray}
\noindent where $\mathbf{r}_i$ are the coordinates of electron $i$, $\mathbf{R}_\alpha$ are the coordinates of nucleus $\alpha$ both expressed in a Cartesian coordinate system whose origin is the center of the positive charge of the \hhh\ system. In eq.~(\ref{dipole}) we use the fact that the electronic dipole-moment operator, $-\sum_{i=1}^2 \mathbf{r}_i$, is
symmetric with respect to the permutation of the electrons and ${\cal P} = {\cal P} {\cal P}$. The spatial wave function is assumed to be normalized. The dipole moment matrix element involving two ECGs, $ g_k(\mathbf{r})$ and $g_l(\mathbf{r})$ is given as: \cite{cafi_thesis}
\begin{equation}
\label{eq:dip2}
\langle g_k(\mathbf{r})| \mathbf{r}_i | g_l(\mathbf{r}) \rangle = S_{kl}
 \bigg\{\left[ \overline{\mathbf{A}}_k + \overline{\mathbf{A}}_l \right]^{-1} 
 \left[ \overline{\mathbf{A}}_k \mathbf{s}_k + \overline{\mathbf{A}}_l 
 \mathbf{s}_l \right]\bigg\}_i,
\end{equation}
\noindent where $\{\mathbf{v}\}_i$ denotes the $i$-th component of vector $\mathbf{v}$, $S_{kl}=\langle g_k | g_l\rangle$
is the overlap integral. 

Since the fits to produce a DMS were not entirely stable we only used dipole components from 4257 points lying up to 30~000 cm$^{-1}$ above the minimum generated  by the ECG calculations in the form $\mu_x$ and $\mu_y$. Previous \hhh\ DMSs were based on a grid of only 69 points; following these studies, \cite{mbb86,Rohse1994}  the dipole moments were transferred to effective charges $d_x$ and $d_y$ which were expanded into a power series in terms of symmetry coordinates. Since our DMS showed sensitivity to the fitting procedure, we opted for the simplest form and used only 7 constants and the function of Ref.\,\citenum{Rohse1994} which reproduce the computed dipoles for points lying below 18\,000 cm$^{-1}$ with a standard deviation of 0.001 a.u. Appropriate functional forms for fits of a global DMS will be the subject of further work. Transition intensities were computed using nuclear motion program DVR3D, \cite{jt338} the non-adiabatic model of Polyansky and Tennyson \cite{jt236} with other parameters chosen as in our previous studies of \hhh\ VIS transition frequencies. \cite{jt512,jt526} Since the major uncertainty in the calculations is given by the fit of the DMS, we used the extremes of plausible fits to estimate uncertainties in our calculated intensities; these uncertainties (see Table 1) are systematic rather than statistical errors and roughly represent a 1-$\sigma$ confidence level.

The high agreement between experiment and the present calculations is shown in Table \ref{tab1:intensities} and Fig.\ \ref{fig1:intensities} and covers the entire range of $B$ coefficients whose magnitude varies by a factor of 300. For comparison, the figure also shows the results of Neale, Miller and Tennyson \cite{jt181} (NMT), the only previous intensity calculation extending into the VIS. The NMT results also agree well with the observations which is somewhat surprising given that the study was not designed to be accurate above the barrier to linearity and that NMT's frequencies are significantly less accurate than the ECG ones. \cite{jt512}

\hhh\ spectra provide a unique probe of the Universe but their interpretation relies heavily on the reliability of {\it ab initio}
intensity calculations. The present studies suggest that, with care, the intensities can be computed over a wide wavelength range. Following the confirmation of the accuracy of our {\it ab initio} \hhh\ model for both transition frequencies \cite{jt512} and, with the present study, intensities, work on a new, experimentally tested, comprehensive \hhh\ line list for astronomical and other applications is now in progress. This will make highly excited levels of \hhh\ broadly available as a tool of precision laser spectroscopy, enabling ultra-sensitive spectroscopic detection of \hhh\ by multistep photodissociation. \cite{Petrignani:2010} It will also pave the way to the reliable modeling of broadband spectral emission from \hhh\ in astrophysical environments, as considered regarding the formation \cite{06GlSa} and evolution \cite{jt327} of metal-poor stars, and the stability of the atmospheres of hot exoplanets. \cite{koskinen:2007} Furthermore, the use of accurate intensities for interpreting the Carrington-Kennedy near-dissociation \hhh\ will allow the division of the spectra in hierarchies, greatly simplifying its analysis. \cite{jt200} 


\begin{acknowledgments}
We thank the Russian Fund for Fundamental Studies, the European Research Council (ERC Advanced Investigator Project 267219), and the Max Planck Society for support.
\end{acknowledgments}


\begin{thebibliography}{32}%
\makeatletter
\providecommand \@ifxundefined [1]{%
 \@ifx{#1\undefined}
}%
\providecommand \@ifnum [1]{%
 \ifnum #1\expandafter \@firstoftwo
 \else \expandafter \@secondoftwo
 \fi
}%
\providecommand \@ifx [1]{%
 \ifx #1\expandafter \@firstoftwo
 \else \expandafter \@secondoftwo
 \fi
}%
\providecommand \natexlab [1]{#1}%
\providecommand \enquote  [1]{``#1''}%
\providecommand \bibnamefont  [1]{#1}%
\providecommand \bibfnamefont [1]{#1}%
\providecommand \citenamefont [1]{#1}%
\providecommand \href@noop [0]{\@secondoftwo}%
\providecommand \href [0]{\begingroup \@sanitize@url \@href}%
\providecommand \@href[1]{\@@startlink{#1}\@@href}%
\providecommand \@@href[1]{\endgroup#1\@@endlink}%
\providecommand \@sanitize@url [0]{\catcode `\\12\catcode `\$12\catcode
  `\&12\catcode `\#12\catcode `\^12\catcode `\_12\catcode `\%12\relax}%
\providecommand \@@startlink[1]{}%
\providecommand \@@endlink[0]{}%
\providecommand \url  [0]{\begingroup\@sanitize@url \@url }%
\providecommand \@url [1]{\endgroup\@href {#1}{\urlprefix }}%
\providecommand \urlprefix  [0]{URL }%
\providecommand \Eprint [0]{\href }%
\providecommand \doibase [0]{http://dx.doi.org/}%
\providecommand \selectlanguage [0]{\@gobble}%
\providecommand \bibinfo  [0]{\@secondoftwo}%
\providecommand \bibfield  [0]{\@secondoftwo}%
\providecommand \translation [1]{[#1]}%
\providecommand \BibitemOpen [0]{}%
\providecommand \bibitemStop [0]{}%
\providecommand \bibitemNoStop [0]{.\EOS\space}%
\providecommand \EOS [0]{\spacefactor3000\relax}%
\providecommand \BibitemShut  [1]{\csname bibitem#1\endcsname}%
\let\auto@bib@innerbib\@empty
\bibitem [{\citenamefont {Tennyson}(1995)}]{jt157}%
  \BibitemOpen
  \bibfield  {author} {\bibinfo {author} {\bibfnamefont {J.}~\bibnamefont
  {Tennyson}},\ }\href@noop {} {\bibfield  {journal} {\bibinfo  {journal} {Rep.
  Prog. Phys.}\ }\textbf {\bibinfo {volume} {58}},\ \bibinfo {pages} {421}
  (\bibinfo {year} {1995})}\BibitemShut {NoStop}%
\bibitem [{\citenamefont {McCall}\ \emph {et~al.}(1999)\citenamefont {McCall},
  \citenamefont {Geballe}, \citenamefont {Hinkle},\ and\ \citenamefont
  {Oka}}]{mccall:1999}%
  \BibitemOpen
  \bibfield  {author} {\bibinfo {author} {\bibfnamefont {B.~J.}\ \bibnamefont
  {McCall}}, \bibinfo {author} {\bibfnamefont {T.~R.}\ \bibnamefont {Geballe}},
  \bibinfo {author} {\bibfnamefont {K.~H.}\ \bibnamefont {Hinkle}}, \ and\
  \bibinfo {author} {\bibfnamefont {T.}~\bibnamefont {Oka}},\ }\href@noop {}
  {\bibfield  {journal} {\bibinfo  {journal} {Astrophys. J.}\ }\textbf
  {\bibinfo {volume} {522}},\ \bibinfo {pages} {338} (\bibinfo {year}
  {1999})}\BibitemShut {NoStop}%
\bibitem [{\citenamefont {Indriolo}\ and\ \citenamefont
  {McCall}(2012)}]{12InMc}%
  \BibitemOpen
  \bibfield  {author} {\bibinfo {author} {\bibfnamefont {N.}~\bibnamefont
  {Indriolo}}\ and\ \bibinfo {author} {\bibfnamefont {B.~J.}\ \bibnamefont
  {McCall}},\ }\href {\doibase {10.1088/0004-637X/745/1/91}} {\bibfield
  {journal} {\bibinfo  {journal} {Astrophys. J.}\ }\textbf {\bibinfo {volume}
  {{745}}},\ \bibinfo {pages} {{91}} (\bibinfo {year} {{2012}})}\BibitemShut
  {NoStop}%
\bibitem [{\citenamefont {Miller}\ \emph {et~al.}(2000)\citenamefont {Miller},
  \citenamefont {Achilleos}, \citenamefont {Ballester}, \citenamefont
  {Geballe}, \citenamefont {Joseph}, \citenamefont {Prange}, \citenamefont
  {Rego}, \citenamefont {Stallard}, \citenamefont {Tennyson}, \citenamefont
  {Trafton},\ and\ \citenamefont {{Waite Jr}}}]{jt258}%
  \BibitemOpen
  \bibfield  {author} {\bibinfo {author} {\bibfnamefont {S.}~\bibnamefont
  {Miller}}, \bibinfo {author} {\bibfnamefont {N.}~\bibnamefont {Achilleos}},
  \bibinfo {author} {\bibfnamefont {G.~E.}\ \bibnamefont {Ballester}}, \bibinfo
  {author} {\bibfnamefont {T.~R.}\ \bibnamefont {Geballe}}, \bibinfo {author}
  {\bibfnamefont {R.~D.}\ \bibnamefont {Joseph}}, \bibinfo {author}
  {\bibfnamefont {R.}~\bibnamefont {Prange}}, \bibinfo {author} {\bibfnamefont
  {D.}~\bibnamefont {Rego}}, \bibinfo {author} {\bibfnamefont {T.}~\bibnamefont
  {Stallard}}, \bibinfo {author} {\bibfnamefont {J.}~\bibnamefont {Tennyson}},
  \bibinfo {author} {\bibfnamefont {L.~M.}\ \bibnamefont {Trafton}}, \ and\
  \bibinfo {author} {\bibfnamefont {J.~H.}\ \bibnamefont {{Waite Jr}}},\
  }\href@noop {} {\bibfield  {journal} {\bibinfo  {journal} {Phil. Trans. Royal
  Soc. London A}\ }\textbf {\bibinfo {volume} {358}},\ \bibinfo {pages} {2485}
  (\bibinfo {year} {2000})}\BibitemShut {NoStop}%
\bibitem [{\citenamefont {Koskinen}\ \emph {et~al.}(2007)\citenamefont
  {Koskinen}, \citenamefont {Aylward},\ and\ \citenamefont
  {Miller}}]{koskinen:2007}%
  \BibitemOpen
  \bibfield  {author} {\bibinfo {author} {\bibfnamefont {T.~T.}\ \bibnamefont
  {Koskinen}}, \bibinfo {author} {\bibfnamefont {A.~D.}\ \bibnamefont
  {Aylward}}, \ and\ \bibinfo {author} {\bibfnamefont {S.}~\bibnamefont
  {Miller}},\ }\href@noop {} {\bibfield  {journal} {\bibinfo  {journal}
  {Nature}\ }\textbf {\bibinfo {volume} {450}},\ \bibinfo {pages} {845}
  (\bibinfo {year} {2007})}\BibitemShut {NoStop}%
\bibitem [{\citenamefont {Kao}\ \emph {et~al.}(1991)\citenamefont {Kao},
  \citenamefont {Oka}, \citenamefont {Miller},\ and\ \citenamefont
  {Tennyson}}]{jt107}%
  \BibitemOpen
  \bibfield  {author} {\bibinfo {author} {\bibfnamefont {L.}~\bibnamefont
  {Kao}}, \bibinfo {author} {\bibfnamefont {T.}~\bibnamefont {Oka}}, \bibinfo
  {author} {\bibfnamefont {S.}~\bibnamefont {Miller}}, \ and\ \bibinfo {author}
  {\bibfnamefont {J.}~\bibnamefont {Tennyson}},\ }\href@noop {} {\bibfield
  {journal} {\bibinfo  {journal} {Astrophys. J. Suppl.}\ }\textbf {\bibinfo
  {volume} {77}},\ \bibinfo {pages} {317} (\bibinfo {year} {1991})}\BibitemShut
  {NoStop}%
\bibitem [{\citenamefont {Neale}\ \emph {et~al.}(1996)\citenamefont {Neale},
  \citenamefont {Miller},\ and\ \citenamefont {Tennyson}}]{jt181}%
  \BibitemOpen
  \bibfield  {author} {\bibinfo {author} {\bibfnamefont {L.}~\bibnamefont
  {Neale}}, \bibinfo {author} {\bibfnamefont {S.}~\bibnamefont {Miller}}, \
  and\ \bibinfo {author} {\bibfnamefont {J.}~\bibnamefont {Tennyson}},\
  }\href@noop {} {\bibfield  {journal} {\bibinfo  {journal} {Astrophys. J.}\
  }\textbf {\bibinfo {volume} {464}},\ \bibinfo {pages} {516} (\bibinfo {year}
  {1996})}\BibitemShut {NoStop}%
\bibitem [{\citenamefont {Lindsay}\ and\ \citenamefont
  {McCall}(2001)}]{lindsay:2001a}%
  \BibitemOpen
  \bibfield  {author} {\bibinfo {author} {\bibfnamefont {C.~M.}\ \bibnamefont
  {Lindsay}}\ and\ \bibinfo {author} {\bibfnamefont {B.~J.}\ \bibnamefont
  {McCall}},\ }\href@noop {} {\bibfield  {journal} {\bibinfo  {journal} {J.
  Mol. Spectrosc.}\ }\textbf {\bibinfo {volume} {210}},\ \bibinfo {pages}
  {60�} (\bibinfo {year} {2001})}\BibitemShut {NoStop}%
\bibitem [{\citenamefont {Pavanello}\ \emph
  {et~al.}(2012{\natexlab{a}})\citenamefont {Pavanello}, \citenamefont
  {Adamowicz}, \citenamefont {Alijah}, \citenamefont {Zobov}, \citenamefont
  {Mizus}, \citenamefont {Polyansky}, \citenamefont {Tennyson}, \citenamefont
  {Szidarovszky}, \citenamefont {Cs\'asz\'ar}, \citenamefont {Berg},
  \citenamefont {Petrignani},\ and\ \citenamefont {Wolf}}]{jt512}%
  \BibitemOpen
  \bibfield  {author} {\bibinfo {author} {\bibfnamefont {M.}~\bibnamefont
  {Pavanello}}, \bibinfo {author} {\bibfnamefont {L.}~\bibnamefont
  {Adamowicz}}, \bibinfo {author} {\bibfnamefont {A.}~\bibnamefont {Alijah}},
  \bibinfo {author} {\bibfnamefont {N.~F.}\ \bibnamefont {Zobov}}, \bibinfo
  {author} {\bibfnamefont {I.~I.}\ \bibnamefont {Mizus}}, \bibinfo {author}
  {\bibfnamefont {O.~L.}\ \bibnamefont {Polyansky}}, \bibinfo {author}
  {\bibfnamefont {J.}~\bibnamefont {Tennyson}}, \bibinfo {author}
  {\bibfnamefont {T.}~\bibnamefont {Szidarovszky}}, \bibinfo {author}
  {\bibfnamefont {A.~G.}\ \bibnamefont {Cs\'asz\'ar}}, \bibinfo {author}
  {\bibfnamefont {M.}~\bibnamefont {Berg}}, \bibinfo {author} {\bibfnamefont
  {A.}~\bibnamefont {Petrignani}}, \ and\ \bibinfo {author} {\bibfnamefont
  {A.}~\bibnamefont {Wolf}},\ }\href@noop {} {\bibfield  {journal} {\bibinfo
  {journal} {Phys. Rev. Lett.}\ }\textbf {\bibinfo {volume} {108}},\ \bibinfo
  {pages} {023002} (\bibinfo {year} {2012}{\natexlab{a}})}\BibitemShut
  {NoStop}%
\bibitem [{\citenamefont {Morong}\ \emph {et~al.}(2009)\citenamefont {Morong},
  \citenamefont {Gottfried},\ and\ \citenamefont {Oka}}]{Morong:2009}%
  \BibitemOpen
  \bibfield  {author} {\bibinfo {author} {\bibfnamefont {C.~P.}\ \bibnamefont
  {Morong}}, \bibinfo {author} {\bibfnamefont {J.~L.}\ \bibnamefont
  {Gottfried}}, \ and\ \bibinfo {author} {\bibfnamefont {T.}~\bibnamefont
  {Oka}},\ }\href@noop {} {\bibfield  {journal} {\bibinfo  {journal} {J. Mol.
  Spectrosc.}\ }\textbf {\bibinfo {volume} {255}},\ \bibinfo {pages} {13}
  (\bibinfo {year} {2009})}\BibitemShut {NoStop}%
\bibitem [{\citenamefont {Lodi}\ \emph {et~al.}(2014)\citenamefont {Lodi},
  \citenamefont {Polyansky}, \citenamefont {Tennyson}, \citenamefont {Alijah},\
  and\ \citenamefont {Zobov}}]{jt582}%
  \BibitemOpen
  \bibfield  {author} {\bibinfo {author} {\bibfnamefont {L.}~\bibnamefont
  {Lodi}}, \bibinfo {author} {\bibfnamefont {O.~L.}\ \bibnamefont {Polyansky}},
  \bibinfo {author} {\bibfnamefont {J.}~\bibnamefont {Tennyson}}, \bibinfo
  {author} {\bibfnamefont {A.}~\bibnamefont {Alijah}}, \ and\ \bibinfo {author}
  {\bibfnamefont {N.~F.}\ \bibnamefont {Zobov}},\ }\href@noop {} {\bibfield
  {journal} {\bibinfo  {journal} {Phys. Rev. A}\ }\textbf {\bibinfo {volume}
  {89}},\ \bibinfo {pages} {032505} (\bibinfo {year} {2014})}\BibitemShut
  {NoStop}%
\bibitem [{\citenamefont {Diniz}\ \emph {et~al.}(2013)\citenamefont {Diniz},
  \citenamefont {Mohallem}, \citenamefont {Alijah}, \citenamefont {Pavanello},
  \citenamefont {Adamowicz}, \citenamefont {Polyansky},\ and\ \citenamefont
  {Tennyson}}]{jt566}%
  \BibitemOpen
  \bibfield  {author} {\bibinfo {author} {\bibfnamefont {L.~G.}\ \bibnamefont
  {Diniz}}, \bibinfo {author} {\bibfnamefont {J.~R.}\ \bibnamefont {Mohallem}},
  \bibinfo {author} {\bibfnamefont {A.}~\bibnamefont {Alijah}}, \bibinfo
  {author} {\bibfnamefont {M.}~\bibnamefont {Pavanello}}, \bibinfo {author}
  {\bibfnamefont {L.}~\bibnamefont {Adamowicz}}, \bibinfo {author}
  {\bibfnamefont {O.~L.}\ \bibnamefont {Polyansky}}, \ and\ \bibinfo {author}
  {\bibfnamefont {J.}~\bibnamefont {Tennyson}},\ }\href {\doibase
  10.1103/PhysRevA.88.032506} {\bibfield  {journal} {\bibinfo  {journal} {Phys.
  Rev. A}\ }\textbf {\bibinfo {volume} {88}},\ \bibinfo {pages} {032506}
  (\bibinfo {year} {2013})}\BibitemShut {NoStop}%
\bibitem [{\citenamefont {Watson}(1994)}]{Watson:1994}%
  \BibitemOpen
  \bibfield  {author} {\bibinfo {author} {\bibfnamefont {J.~K.~G.}\
  \bibnamefont {Watson}},\ }\href {\doibase 10.1139/p94-092} {\bibfield
  {journal} {\bibinfo  {journal} {Can. J. Phys.}\ }\textbf {\bibinfo {volume}
  {72}},\ \bibinfo {pages} {702} (\bibinfo {year} {1994})}\BibitemShut
  {NoStop}%
\bibitem [{\citenamefont {Polyansky}\ and\ \citenamefont
  {Tennyson}(1999)}]{jt236}%
  \BibitemOpen
  \bibfield  {author} {\bibinfo {author} {\bibfnamefont {O.~L.}\ \bibnamefont
  {Polyansky}}\ and\ \bibinfo {author} {\bibfnamefont {J.}~\bibnamefont
  {Tennyson}},\ }\href@noop {} {\bibfield  {journal} {\bibinfo  {journal} {J.
  Chem. Phys.}\ }\textbf {\bibinfo {volume} {110}},\ \bibinfo {pages} {5056}
  (\bibinfo {year} {1999})}\BibitemShut {NoStop}%
\bibitem [{\citenamefont {Carrington}\ \emph {et~al.}(1982)\citenamefont
  {Carrington}, \citenamefont {Buttenshaw},\ and\ \citenamefont
  {Kennedy}}]{Carrington:1982}%
  \BibitemOpen
  \bibfield  {author} {\bibinfo {author} {\bibfnamefont {A.}~\bibnamefont
  {Carrington}}, \bibinfo {author} {\bibfnamefont {J.}~\bibnamefont
  {Buttenshaw}}, \ and\ \bibinfo {author} {\bibfnamefont {R.}~\bibnamefont
  {Kennedy}},\ }\href@noop {} {\bibfield  {journal} {\bibinfo  {journal} {Mol.
  Phys.}\ }\textbf {\bibinfo {volume} {45}},\ \bibinfo {pages} {753} (\bibinfo
  {year} {1982})}\BibitemShut {NoStop}%
\bibitem [{\citenamefont {Carrington}\ and\ \citenamefont
  {Kennedy}(1984)}]{Carrington:1984}%
  \BibitemOpen
  \bibfield  {author} {\bibinfo {author} {\bibfnamefont {A.}~\bibnamefont
  {Carrington}}\ and\ \bibinfo {author} {\bibfnamefont {R.~A.}\ \bibnamefont
  {Kennedy}},\ }\href@noop {} {\bibfield  {journal} {\bibinfo  {journal} {J.
  Chem. Phys.}\ }\textbf {\bibinfo {volume} {81}},\ \bibinfo {pages} {91}
  (\bibinfo {year} {1984})}\BibitemShut {NoStop}%
\bibitem [{\citenamefont {Gottfried}(2006)}]{Gottfried:2006}%
  \BibitemOpen
  \bibfield  {author} {\bibinfo {author} {\bibfnamefont {J.~L.}\ \bibnamefont
  {Gottfried}},\ }\href@noop {} {\bibfield  {journal} {\bibinfo  {journal}
  {Phil. Trans. R. Soc. A}\ }\textbf {\bibinfo {volume} {364}},\ \bibinfo
  {pages} {2917�} (\bibinfo {year} {2006})}\BibitemShut {NoStop}%
\bibitem [{\citenamefont {Gottfried}\ \emph {et~al.}(2003)\citenamefont
  {Gottfried}, \citenamefont {McCall},\ and\ \citenamefont
  {Oka}}]{Gottfried:2003}%
  \BibitemOpen
  \bibfield  {author} {\bibinfo {author} {\bibfnamefont {J.~L.}\ \bibnamefont
  {Gottfried}}, \bibinfo {author} {\bibfnamefont {B.~J.}\ \bibnamefont
  {McCall}}, \ and\ \bibinfo {author} {\bibfnamefont {T.}~\bibnamefont {Oka}},\
  }\href@noop {} {\bibfield  {journal} {\bibinfo  {journal} {J. Chem. Phys.}\
  }\textbf {\bibinfo {volume} {118}},\ \bibinfo {pages} {10890} (\bibinfo
  {year} {2003})}\BibitemShut {NoStop}%
\bibitem [{\citenamefont {Kreckel}\ \emph {et~al.}(2008)\citenamefont
  {Kreckel}, \citenamefont {Bing}, \citenamefont {Reinhardt}, \citenamefont
  {Petrignani}, \citenamefont {Berg},\ and\ \citenamefont
  {Wolf}}]{Kreckel:2008}%
  \BibitemOpen
  \bibfield  {author} {\bibinfo {author} {\bibfnamefont {H.}~\bibnamefont
  {Kreckel}}, \bibinfo {author} {\bibfnamefont {D.}~\bibnamefont {Bing}},
  \bibinfo {author} {\bibfnamefont {S.}~\bibnamefont {Reinhardt}}, \bibinfo
  {author} {\bibfnamefont {A.}~\bibnamefont {Petrignani}}, \bibinfo {author}
  {\bibfnamefont {M.~H.}\ \bibnamefont {Berg}}, \ and\ \bibinfo {author}
  {\bibfnamefont {A.}~\bibnamefont {Wolf}},\ }\href@noop {} {\bibfield
  {journal} {\bibinfo  {journal} {J. Chem. Phys.}\ }\textbf {\bibinfo {volume}
  {129}},\ \bibinfo {pages} {164312} (\bibinfo {year} {2008})}\BibitemShut
  {NoStop}%
\bibitem [{\citenamefont {Berg}\ \emph {et~al.}(2012)\citenamefont {Berg},
  \citenamefont {Wolf},\ and\ \citenamefont {Petrignani}}]{Berg:2012}%
  \BibitemOpen
  \bibfield  {author} {\bibinfo {author} {\bibfnamefont {M.}~\bibnamefont
  {Berg}}, \bibinfo {author} {\bibfnamefont {A.}~\bibnamefont {Wolf}}, \ and\
  \bibinfo {author} {\bibfnamefont {A.}~\bibnamefont {Petrignani}},\
  }\href@noop {} {\bibfield  {journal} {\bibinfo  {journal} {Phil. Trans. R.
  Soc. A}\ ,\ \bibinfo {pages} {5028}} (\bibinfo {year} {2012})}\BibitemShut
  {NoStop}%
\bibitem [{\citenamefont {Daly}(1960)}]{Daly:1960}%
  \BibitemOpen
  \bibfield  {author} {\bibinfo {author} {\bibfnamefont {N.~R.}\ \bibnamefont
  {Daly}},\ }\href@noop {} {\bibfield  {journal} {\bibinfo  {journal} {Rev.
  Sci. Instrum.}\ }\textbf {\bibinfo {volume} {31}},\ \bibinfo {pages} {264}
  (\bibinfo {year} {1960})}\BibitemShut {NoStop}%
\bibitem [{\citenamefont {Alijah}(2010)}]{alijah10}%
  \BibitemOpen
  \bibfield  {author} {\bibinfo {author} {\bibfnamefont {A.}~\bibnamefont
  {Alijah}},\ }\href {\doibase 10.1016/j.jms.2010.09.009} {\bibfield  {journal}
  {\bibinfo  {journal} {J. Mol. Spectrosc.}\ }\textbf {\bibinfo {volume}
  {264}},\ \bibinfo {pages} {111} (\bibinfo {year} {2010})}\BibitemShut
  {NoStop}%
\bibitem [{\citenamefont {Pavanello}\ \emph
  {et~al.}(2012{\natexlab{b}})\citenamefont {Pavanello}, \citenamefont
  {Adamowicz}, \citenamefont {Alijah}, \citenamefont {Zobov}, \citenamefont
  {Mizus}, \citenamefont {Polyansky}, \citenamefont {Tennyson}, \citenamefont
  {Szidarovszky},\ and\ \citenamefont {Cs\'asz\'ar}}]{jt526}%
  \BibitemOpen
  \bibfield  {author} {\bibinfo {author} {\bibfnamefont {M.}~\bibnamefont
  {Pavanello}}, \bibinfo {author} {\bibfnamefont {L.}~\bibnamefont
  {Adamowicz}}, \bibinfo {author} {\bibfnamefont {A.}~\bibnamefont {Alijah}},
  \bibinfo {author} {\bibfnamefont {N.~F.}\ \bibnamefont {Zobov}}, \bibinfo
  {author} {\bibfnamefont {I.~I.}\ \bibnamefont {Mizus}}, \bibinfo {author}
  {\bibfnamefont {O.~L.}\ \bibnamefont {Polyansky}}, \bibinfo {author}
  {\bibfnamefont {J.}~\bibnamefont {Tennyson}}, \bibinfo {author}
  {\bibfnamefont {T.}~\bibnamefont {Szidarovszky}}, \ and\ \bibinfo {author}
  {\bibfnamefont {A.~G.}\ \bibnamefont {Cs\'asz\'ar}},\ }\href@noop {}
  {\bibfield  {journal} {\bibinfo  {journal} {J. Chem. Phys.}\ }\textbf
  {\bibinfo {volume} {136}},\ \bibinfo {pages} {184303} (\bibinfo {year}
  {2012}{\natexlab{b}})}\BibitemShut {NoStop}%
\bibitem [{\citenamefont {Bubin}\ \emph {et~al.}(2012)\citenamefont {Bubin},
  \citenamefont {Pavanello}, \citenamefont {Tung}, \citenamefont {Sharkey},\
  and\ \citenamefont {Adamowicz}}]{bubin2012a}%
  \BibitemOpen
  \bibfield  {author} {\bibinfo {author} {\bibfnamefont {S.}~\bibnamefont
  {Bubin}}, \bibinfo {author} {\bibfnamefont {M.}~\bibnamefont {Pavanello}},
  \bibinfo {author} {\bibfnamefont {W.-C.}\ \bibnamefont {Tung}}, \bibinfo
  {author} {\bibfnamefont {K.~L.}\ \bibnamefont {Sharkey}}, \ and\ \bibinfo
  {author} {\bibfnamefont {L.}~\bibnamefont {Adamowicz}},\ }\href {\doibase
  10.1021/cr200419d} {\bibfield  {journal} {\bibinfo  {journal} {Chem. Rev.}\
  }\textbf {\bibinfo {volume} {113}},\ \bibinfo {pages} {36} (\bibinfo {year}
  {2012})}\BibitemShut {NoStop}%
\bibitem [{\citenamefont {Cafiero}(2002)}]{cafi_thesis}%
  \BibitemOpen
  \bibfield  {author} {\bibinfo {author} {\bibfnamefont {M.~L.}\ \bibnamefont
  {Cafiero}},\ }\emph {\bibinfo {title} {{High accuracy calculations on
  Coulombic few particle systems in a basis of explicitly correlated Gaussian
  functions}}},\ \href {http://hdl.handle.net/10150/280156} {Ph.D. thesis},\
  \bibinfo  {school} {{The University of Arizona}} (\bibinfo {year}
  {2002})\BibitemShut {NoStop}%
\bibitem [{\citenamefont {Meyer}\ \emph {et~al.}(1986)\citenamefont {Meyer},
  \citenamefont {Botschwina},\ and\ \citenamefont {Burton}}]{mbb86}%
  \BibitemOpen
  \bibfield  {author} {\bibinfo {author} {\bibfnamefont {W.}~\bibnamefont
  {Meyer}}, \bibinfo {author} {\bibfnamefont {P.}~\bibnamefont {Botschwina}}, \
  and\ \bibinfo {author} {\bibfnamefont {P.~G.}\ \bibnamefont {Burton}},\
  }\href@noop {} {\bibfield  {journal} {\bibinfo  {journal} {Chem. Phys.}\
  }\textbf {\bibinfo {volume} {82}},\ \bibinfo {pages} {891} (\bibinfo {year}
  {1986})}\BibitemShut {NoStop}%
\bibitem [{\citenamefont {R\"{o}hse}\ \emph {et~al.}(1994)\citenamefont
  {R\"{o}hse}, \citenamefont {Kutzelnigg}, \citenamefont {Jaquet},\ and\
  \citenamefont {Klopper}}]{Rohse1994}%
  \BibitemOpen
  \bibfield  {author} {\bibinfo {author} {\bibfnamefont {R.}~\bibnamefont
  {R\"{o}hse}}, \bibinfo {author} {\bibfnamefont {W.}~\bibnamefont
  {Kutzelnigg}}, \bibinfo {author} {\bibfnamefont {R.}~\bibnamefont {Jaquet}},
  \ and\ \bibinfo {author} {\bibfnamefont {W.}~\bibnamefont {Klopper}},\
  }\href@noop {} {\bibfield  {journal} {\bibinfo  {journal} {J. Chem. Phys.}\
  }\textbf {\bibinfo {volume} {101}},\ \bibinfo {pages} {2231} (\bibinfo {year}
  {1994})}\BibitemShut {NoStop}%
\bibitem [{\citenamefont {Tennyson}\ \emph {et~al.}(2004)\citenamefont
  {Tennyson}, \citenamefont {Kostin}, \citenamefont {Barletta}, \citenamefont
  {Harris}, \citenamefont {Polyansky}, \citenamefont {Ramanlal},\ and\
  \citenamefont {Zobov}}]{jt338}%
  \BibitemOpen
  \bibfield  {author} {\bibinfo {author} {\bibfnamefont {J.}~\bibnamefont
  {Tennyson}}, \bibinfo {author} {\bibfnamefont {M.~A.}\ \bibnamefont
  {Kostin}}, \bibinfo {author} {\bibfnamefont {P.}~\bibnamefont {Barletta}},
  \bibinfo {author} {\bibfnamefont {G.~J.}\ \bibnamefont {Harris}}, \bibinfo
  {author} {\bibfnamefont {O.~L.}\ \bibnamefont {Polyansky}}, \bibinfo {author}
  {\bibfnamefont {J.}~\bibnamefont {Ramanlal}}, \ and\ \bibinfo {author}
  {\bibfnamefont {N.~F.}\ \bibnamefont {Zobov}},\ }\href@noop {} {\bibfield
  {journal} {\bibinfo  {journal} {Comput. Phys. Commun.}\ }\textbf {\bibinfo
  {volume} {163}},\ \bibinfo {pages} {85} (\bibinfo {year} {2004})}\BibitemShut
  {NoStop}%
\bibitem [{\citenamefont {Petrignani}\ \emph {et~al.}(2010)\citenamefont
  {Petrignani}, \citenamefont {Bing}, \citenamefont {Novotn\'{y}},
  \citenamefont {Berg}, \citenamefont {Buhr}, \citenamefont {Grieser},
  \citenamefont {Jordon-Thaden}, \citenamefont {Krantz}, \citenamefont
  {Mendes}, \citenamefont {Menk}, \citenamefont {Novotny}, \citenamefont
  {Orlov}, \citenamefont {Repnow}, \citenamefont {St\"{u}tzel}, \citenamefont
  {Urbain},\ and\ \citenamefont {Wolf}}]{Petrignani:2010}%
  \BibitemOpen
  \bibfield  {author} {\bibinfo {author} {\bibfnamefont {A.}~\bibnamefont
  {Petrignani}}, \bibinfo {author} {\bibfnamefont {D.}~\bibnamefont {Bing}},
  \bibinfo {author} {\bibfnamefont {O.}~\bibnamefont {Novotn\'{y}}}, \bibinfo
  {author} {\bibfnamefont {M.~H.}\ \bibnamefont {Berg}}, \bibinfo {author}
  {\bibfnamefont {H.}~\bibnamefont {Buhr}}, \bibinfo {author} {\bibfnamefont
  {M.}~\bibnamefont {Grieser}}, \bibinfo {author} {\bibfnamefont
  {B.}~\bibnamefont {Jordon-Thaden}}, \bibinfo {author} {\bibfnamefont
  {C.}~\bibnamefont {Krantz}}, \bibinfo {author} {\bibfnamefont {M.~B.}\
  \bibnamefont {Mendes}}, \bibinfo {author} {\bibfnamefont {S.}~\bibnamefont
  {Menk}}, \bibinfo {author} {\bibfnamefont {S.}~\bibnamefont {Novotny}},
  \bibinfo {author} {\bibfnamefont {D.~A.}\ \bibnamefont {Orlov}}, \bibinfo
  {author} {\bibfnamefont {R.}~\bibnamefont {Repnow}}, \bibinfo {author}
  {\bibfnamefont {J.}~\bibnamefont {St\"{u}tzel}}, \bibinfo {author}
  {\bibfnamefont {X.}~\bibnamefont {Urbain}}, \ and\ \bibinfo {author}
  {\bibfnamefont {A.}~\bibnamefont {Wolf}},\ }\href@noop {} {\bibfield
  {journal} {\bibinfo  {journal} {J. Phys. Chem. A}\ }\textbf {\bibinfo
  {volume} {114}},\ \bibinfo {pages} {4864} (\bibinfo {year}
  {2010})}\BibitemShut {NoStop}%
\bibitem [{\citenamefont {Glover}\ and\ \citenamefont {Savin}(2006)}]{06GlSa}%
  \BibitemOpen
  \bibfield  {author} {\bibinfo {author} {\bibfnamefont {S.}~\bibnamefont
  {Glover}}\ and\ \bibinfo {author} {\bibfnamefont {D.~W.}\ \bibnamefont
  {Savin}},\ }\href {\doibase {10.1098/rsta.2006.1867}} {\bibfield  {journal}
  {\bibinfo  {journal} {Phil. Trans. R. Soc. A}\ }\textbf {\bibinfo {volume}
  {{364}}},\ \bibinfo {pages} {3107} (\bibinfo {year} {{2006}})}\BibitemShut
  {NoStop}%
\bibitem [{\citenamefont {Harris}\ \emph {et~al.}(2004)\citenamefont {Harris},
  \citenamefont {Lynas-Gray}, \citenamefont {Miller},\ and\ \citenamefont
  {Tennyson}}]{jt327}%
  \BibitemOpen
  \bibfield  {author} {\bibinfo {author} {\bibfnamefont {G.~J.}\ \bibnamefont
  {Harris}}, \bibinfo {author} {\bibfnamefont {A.~E.}\ \bibnamefont
  {Lynas-Gray}}, \bibinfo {author} {\bibfnamefont {S.}~\bibnamefont {Miller}},
  \ and\ \bibinfo {author} {\bibfnamefont {J.}~\bibnamefont {Tennyson}},\
  }\href@noop {} {\bibfield  {journal} {\bibinfo  {journal} {Astrophys. J.}\
  }\textbf {\bibinfo {volume} {600}},\ \bibinfo {pages} {1025} (\bibinfo {year}
  {2004})}\BibitemShut {NoStop}%
\bibitem [{\citenamefont {Polyansky}\ \emph {et~al.}(1997)\citenamefont
  {Polyansky}, \citenamefont {Zobov}, \citenamefont {Viti}, \citenamefont
  {Tennyson}, \citenamefont {Bernath},\ and\ \citenamefont {Wallace}}]{jt200}%
  \BibitemOpen
  \bibfield  {author} {\bibinfo {author} {\bibfnamefont {O.~L.}\ \bibnamefont
  {Polyansky}}, \bibinfo {author} {\bibfnamefont {N.~F.}\ \bibnamefont
  {Zobov}}, \bibinfo {author} {\bibfnamefont {S.}~\bibnamefont {Viti}},
  \bibinfo {author} {\bibfnamefont {J.}~\bibnamefont {Tennyson}}, \bibinfo
  {author} {\bibfnamefont {P.~F.}\ \bibnamefont {Bernath}}, \ and\ \bibinfo
  {author} {\bibfnamefont {L.}~\bibnamefont {Wallace}},\ }\href@noop {}
  {\bibfield  {journal} {\bibinfo  {journal} {Science}\ }\textbf {\bibinfo
  {volume} {277}},\ \bibinfo {pages} {346} (\bibinfo {year}
  {1997})}\BibitemShut {NoStop}%
\end{thebibliography}

%
\end{document}